# Amorphous carbon a promising material for sodium ion battery anodes: a first principles study

Fleur Legrain, Konstantinos Kotsis, and Sergei Manzhos[a]

Department of Mechanical Engineering, National University of Singapore, Block EA #07-08, 9 Engineering Drive 1, Singapore 117576, Singapore.

**Abstract**

We present a comparative ab initio computational study of sodium and lithium storage in amorphous (glassy) carbon (*a*-C) vs. graphite. Amorphous structures are obtained by fitting stochastically generated structures to a reference radial distribution function. Li insertion is thermodynamically favored in both graphite and *a*-C. While sodium insertion is thermodynamically unfavored in graphite, *a*-C possesses multiple insertion sites with binding energies stronger than Na cohesive energy, making it usable as anode material for Na-ion batteries.

---

[a] Author to whom correspondence should be addressed; E-Mail: mpemanzh@nus.edu.sg;
Tel.: +65-6516-4605; Fax: +65-6779-1459

## 1. Introduction

The perspective of a widespread use of clean but intermittent sources of electricity (wind and solar) as well as of hybrid electric vehicles calls for alternatives to Li-ion batteries, as Li resources may be limited [1, 2]. Sodium being abundant, cheap, and a relatively light and small atom, Na-ion batteries have recently attracted much interest [2, 3]. However, while most studies of Na-ion batteries focus on the positive electrode, the negative electrode remains less investigated and an efficient anode material providing all a good capacity, a high cycle life, and a descent rate of charge/discharge, is still not available. Some efficient electrode materials for Li, in particular diamond Si and graphite C, have been shown to not allow the intercalation of Na [4, 5]. Computational studies report positive intercalation energies [6, 7] and therefore suggest that the insertion of Na into the crystalline framework of graphite and Si is thermodynamically not favored: Na atoms prefer to cluster rather than to intercalate into the crystalline phase. Absence of significant Na insertion in graphite is also confirmed experimentally [8]. Na insertion into graphite has only been achieved by co-insertion with some electrolytes [9].

Amorphization of Si has recently been shown to improve substantially the interaction between Si and Na, and amorphous Si (*a*-Si) has been predicted to allow Na insertion by two independent studies [6, 10]. We hypothesize that the same should hold for amorphization of carbon. To the best of our knowledge, the effect of amorphization of carbon on Na and Li storage has not been computed. Its knowledge is needed for rational design of electrodes for Na-ion batteries as well as for Li-ion batteries. Indeed, carbon is widely used as storage medium and/or conducting binder in both types of batteries [11]. Specifically, nano-sized graphite (which is also sometimes called "amorphous carbon", as opposed to truly amorphous or "glassy" C [12]) has been used as anode for Na-ion batteries [13]. Amorphous carbon as anode material has been investigated for Li insertion [14]. In Ref. [15], amorphous carbon-coated $TiO_2$ nanocrystals were used in a lithium-ion battery, while amorphous carbon-coated $Na_7Fe_7(PO_4)_6F_3$ has recently been investigated experimentally as a cathode material in sodium ion batteries, and it was shown that the coated material exhibits twice the capacity of the uncoated sample [16]. Glassy carbon in nanocomposites can be a component of electrodes. An increasing interest lies in the combination of different materials, composite materials, e.g. amorphous phosphorous/carbon composite [17] and nanocomposite Sb/C [18] were investigated as promising anode materials for Na-ion batteries.

In this paper, we therefore investigate the effect of amorphization of carbon on the energetics of insertion of Li and Na and show that, contrary to graphite, amorphous (glassy) carbon favors sodium insertion and can therefore work as anode for Na-ion batteries. This is a computational study at the dispersion-corrected Density Functional Theory (DFT-D) level. In Section 2, we introduce the methodology used to make the amorphous structures and to do ab initio calculations. In Section 3, we present calculations of two different amorphous carbon structures and the number of the most stable insertion sites into these two structures (Section 3.1) in comparison to the known lowest site in graphite, as well as the energetics of Li and Na in *a*-C versus graphite (Section 3.2). We show that amorphization leads to stronger binding of Li and Na to carbon. Section 4 concludes that amorphous carbon is suitable as anode material Na-ion batteries and may also be advantageous for use in Li-ion batteries.

## 2. Methods

*2.1. Amorphous carbon structure*

Amorphous structures were generated by randomly sampling distributions of 64 carbon atoms placed in a box of size 8x8x8 Å$^3$, with periodic boundary conditions. The initial density of about 2.5 g/cm$^3$ is near that of previously reported amorphous structures [19, 20]. A large number (>10$^6$) of distributions were sampled, for which the radial distribution functions (RDF) were compared to the experimental RDF of Ref. [19]. The structures giving a good fit of the RDF to the experimental RDF [19] were then optimized with DFT-D including optimization of lattice vectors (to zero pressure), which did not result in significant changes of the RDF or of the density. While the method bears similarity to that used in Ref. [12] in that initial random structures are used, we introduce significant improvements in that we fit to the experimental RDF and ensure that the structure is stable under cell vector optimization to target pressure.

*2.2. Ab initio calculations*

Structures were optimized with density functional theory (DFT) [21, 22] using the SIESTA code [23]. The PBE exchange correlation functional [24] and a double-$\zeta$ polarized basis set were used. The basis sets were tuned to reproduce the cohesive energies of Li, Na, and diamond carbon (computed values of, respectivelyt, 1.67, 1.14, and 7.65 eV are in good agreement with reference values [25, 26, 27]). Core electrons were replaced with Troullier-Martins pseudopotentials [28]. Spin-polarized calculations were performed. The DFT-D2 approach of Grimme [29] is used to model the van der Waals interaction between the C atoms. This is important for comparison with graphite. No correction was used between the C and Li/Na atoms due to significant ionicity of the bonds (large charge donation, see below). The Grimme parameters were tuned to reproduce the spacing of layers of graphite (the computed value of 3.35 Å is in good agreement with reference values [30, 31]). This was achieved with Grimme parameter values of $s_6$=1.0, $D$=20, $C_6$= 1.75 J nm$^6$ mol$^{-1}$, $r$=1.725 Å [29]. Nearly cubic supercells of 64 and 128 C atoms with periodic boundary conditions were used to model the intercalation of Li and Na in *a*-C and in graphite, respectively. The Brillouin zone was sampled with 3×3×3 (4×4×4) Monkhorst-Pack point grid [32] for amorphous (graphite) structures, and a 100

Ry cutoff was used for the Fourier expansion of the electron density. All atomic positions and the lattice vectors were allowed to relax, until forces were below 0.03 eV/Å and stresses below 0.1 GPa, respectively.

The energetics of Na and Li insertion in the amorphous and crystalline (graphite) phases of C were analyzed based on the defect formation energies per Li/Na atom $E_f$,

$$E_f = \frac{E(C-M) - E(C) - nE(M)}{n},$$

where M stands for the inserted metal (M=Li/Na), $n$ represents the number of inserted metal atoms of type M, $E$(C-M) designates the energy of the supercell with the Li/Na-inserted C structures, $E$(C) represents the energy of the pure carbon structures (graphite or amorphous), and $E$(M) represents the energy of one atom of Li/Na in bulk (i.e. $bcc$ Li/Na). Negative values of $E_f$ indicate therefore thermodynamical preference for Li/Na insertion, and positive values indicate thermodynamics preference for bulk metal formation ("plating") [33, 34].

## 3. Results and Discussion

*3.1. Amorphous carbon structure and insertion sites*

Two amorphous structures were generated and are shown in Fig. 1(a, b). We use two different structures to ensure that the results are not skewed due to a particular generated structure. Their RDF are shown in Fig. 1(c) together (and in good agreement) with that from neutron diffraction data [19]. Both amorphous structures have a mass density of ~2.5 g/cm$^3$ and a sp3 fraction of ~0.5, in agreement with amorphous structures generated by melting and quenching (MD) using the DFTB scheme [20]. The fractional sp3 character is related to a coordination number of ~3.5 [35, 36]. The amorphous structures are less stable than the graphite phase by 0.80 and 0.92 eV per atom for structures 1 and 2, respectively. To find Li/Na insertion sites in *a*-C, we performed a *k*-means clustering [37, 38] analysis of a uniform three-dimensional grid of points covering each structure spaced by 0.2 Å and excluding points closer than 1.5 Å to C atoms. This allowed us to identify 17 and 19 potential insertion sites in structures 1 and 2, respectively. These sites were used as initial guesses for the insertion sites and further optimized by DFT-D. The positions of optimized unique insertion sites are also shown in Fig. 1 (a, b). A total of 13/12 unique Li/Na sites were found in both structures. The known lowest energy site in graphite is also shown in Fig. 1(d) and is used for comparison [31, 39, 40].

*3.2. Insertion energetics of Li and Na into a-C vs. graphite*

The defect formation energies of Li and Na in all structures are listed in Table 1 and are plotted in Fig. 2(a). The defect formation energy for Li insertion in graphite is -0.09 eV and for Na insertion +0.76, in good agreement with observed anodic voltages for Li [34, 8], with previous calculations [6], and with the fact that Na does not intercalate in graphite [4, 5]. In *a*-C, there is a distribution of $E_f$ values, with lowest $E_f$ values stabilized by 1.48 eV for Li and 1.99 eV for Na, vs. graphite. This behavior is similar to what we previously observed for Li and Na insertion in amorphous vs. crystalline Si and $TiO_2$ [6, 41]. All Li sites except one have a negative $E_f$. More importantly, the amorphization of carbon makes Na insertion thermodynamically favored, with half the sites showing binding energies stronger than the cohesive energy of Na. *a-C will therefore operate as an anode for Na-ion batteries, while graphite does not*. Amorphization of carbon could also be used to increase the anodic voltage in Li-ion batteries by up to 1.5 V, which could be useful e.g. to match it with the redox window of the electrolyte and limit electrolyte decomposition [34, 42].

In Fig. 2(b), we plot $E_f$ as a function of the effective coordination number $N$ (number of neighbors of the Na atom with a cutoff distance $r_c$=2.5 Å). The correlations with Pearson's $R$ values of 0.93 (0.73) for Na (Li) are statistically significant. As expected, the inserted Li and Na atoms donate charge to the host, ranging 0.2-0.6 (0.5-0.7) |$e$| based on Mulliken (Voronoi) charges for Li and 0.4-0.7 (0.4-0.6) |$e$| for Na in *a*-C. To compare, in graphite, the charge donation is 0.5 (0.7) |$e$| for Li and 0.5 (0.6) |$e$| for Na. However, no significant correlation of $E_f$ to the charges was found.

We also investigated the insertion energetics of Li/Na in *a*-C for higher concentrations $x$ ($\frac{2}{64}, \frac{3}{64}, \frac{6}{64}$) than that of $x = \frac{1}{64}$ considered above, $x$ being the number of Li/Na atom inserted per C atom. Because of the prohibitive computational cost of the exhaustive screening of all combinations of two, three, and six occupied sites, only the 2, 3 and 6 lowest energy sites are computed for each amorphous structure (i.e. two calculations for each concentration). The plot of the defect formation energies against Li/Na concentration is given in Fig. 3. The negative defect formation energies for $x = \frac{2}{64}$ and $\frac{3}{64}$ (and also $\frac{6}{64}$ in Li) suggest that the insertion of 2 and 3 Li/Na (and 6 Li) per 64 atoms of C is favored in the amorphous phase. For 6 Na atoms, the configurations modeled give a favored insertion in one of the two amorphous structures, and a slightly positive $E_f$ in the other. The defect formation energies increase with metal concentration, as expected; we note that the negative of the defect formation energy, -$E_f$, is equal to the average voltage, see Ref. [43]. These results suggest that Na intercalation could happen for a higher concentration than the very dilute concentration of $x = \frac{1}{64}$.

## 4. Conclusions

We have shown in a comparative computational study that while the lowest-energy insertion site in graphite does not favor Na intercalation, the amorphous phase (*a*-C) provides insertion sites with a wide distribution of energies, including sites with binding energies of Na stronger than Na cohesive energy. *a-C can therefore operate as an anode for Na-ion batteries, while graphite cannot.*

Insertion of both Li and Na is stabilized by amorphization, by up to 1.5 and 2.0 eV, respectively. Amorphization of carbon could also be used to increase the anodic voltage in Li-ion batteries to e.g. limit electrolyte decomposition.

We also show that it is possible to obtain a reliable amorphous structure in a computationally efficient way by optimizing randomized structures pre-selected to satisfy the desired (e.g. experimental) radial distribution function.

## 5. Acknowledgments

This work was supported by the Ministry of Education of Singapore via an AcRF grant (R-265-000-494-112).

## 6. References

1. J.-M. Tarascon, Is lithium the new gold?, Nature Chemistry 2 (2010) 510.
2. M.D. Slater, D. Kim, E. Lee, C.S. Johnson, Sodium-ion batteries, Adv. Funct. Mater. 23 (2013) 947-958.
3. C. Grosjean, P.H. Miranda, M. Perrin, P. Poggi, Assessment of world lithium resources and consequences of their geographic distribution on the expected development of the electric vehicle industry, Renewable & Sustainable Energy Reviews 16 (2012) 1735-1744.
4. P. Ge, M. Fouletier, Electrochemical intercalation of sodium in graphite, Solid State Ionics 28-30 (1988) 1172-1175.
5. S. Komaba, Y. Matsuura, T. Ishikawa, N. Yabuuchi, W. Murata, S. Kuze, Redox reaction of Sn-polyacrylate electrodes in aprotic Na cell, Electrochem. Commun. 21 (2012) 65-68.
6. F. Legrain, O.I. Malyi, S. Manzhos, Comparative computational study of the energetics of Li, Na, and Mg storage in amorphous and crystalline silicon, Comput. Mater. Sci. 94 (2014) 214-217.
7. Y. Okamoto, Density Functional Theory calculations of alkali metal (Li, Na, and K) graphite intercalation Compounds, J. Phys. Chem. C 118 (2013) 16-19.
8. D. A. Stevens, J.R. Dahn, The mechanisms of lithium and sodium insertion in carbon materials, J. Electrochem. Soc. 148 (2001) A803-A811.
9. H. Kim, J. Hong, Y.-U. Park, J. Kim, I. Hwang, K. Kang, Sodium storage behavior in natural graphite using ether-based electrolyte systems, Adv. Funct. Mater. (2014) DOI: 10.1002/adfm.201402984.
10. S.C. Jung, D.S. Jung, J.W. Choi, Y.-K. Han, Atom-level understanding of the sodiation process in silicon anode material, J. Phys. Chem. Lett. 5 (2014) 1283-88.

11. S.-L. Chou, Y. Pan, J.-Z. Wang, H.-K. Liu, S.-X. Dou, Small things make a big difference: binder effects on the performance of Li and Na batteries, Phys. Chem. Phys. Chem. 16 (2014) 20347-20359.

12. X. Jiang, C. Århammar, P. Liu, J. Zhao, R. Ahuja, The R3-carbon allotrope: a pathway towards glassy carbon under high pressure, Sci. Rep. 3 (2013) DOI:10.1038/srep01877.

13. E.M. Lotfabad, J. Ding, K. Cui, A. Kohandehghan, W.P. Kalisvaart, M. Hazelton, D. Mitlin, High-density sodium and lithium ion battery anodes from banana peels, ACS Nano 8 (2014) 7115-7129.

14. G. Gourdin, P.H. Smith, T. Jiang, T.N. Tran, D. Qu, Lithiation of amorphous carbon negative electrode for Li ion capacitor, J. Electroanal. Chem. 688 (2013) 103-112.

15. T. Xia, W. Zhang, Z. Wang, Y. Zhang, X. Song, J. Murowchick, V. Battaglia, G. Liu, X. Chen, Amorphous carbon-coated $TiO_2$ nanocrystals for improved lithium-ion battery and photocatalytic performance, Nano Energy 6 (2014) 109-118.

16. T. Ramireddy, M.M. Rahman, N. Sharma, A.M. Glushenkov, Y. Chen, Carbon coated $Na_7Fe_7(PO_4)_6F_3$: a novel intercalation cathode for sodium-ion batteries, Journal of Power Sources 271 (2014) 497-503.

17. Y. Kim, Y. Park, A. Choi, N.-S. Choi, J. Kim, J. Lee, J. H. Ryu, S.M. Oh, K.T. Lee, An amorphous red phosphorus/carbon composite as a promising anode material for sodium ion batteries, Adv. Mater. 25 (2013) 3045-3049.

18. J. Qian, Y. Chen, L. Wu, Y. Cao, X. Ai, H. Yang, High capacity Na-storage and superior cyclability of nanocomposite Sb/C anode for Na-ion batteries, Chem. Comm. 48 (2012) 7070-7072.

19. K.W.R. Gilkes, P.H. Gaskell, J. Robertson, Comparison of neutron-scattering data for tetrahedral amorphous carbon with structural models, Phys. Rev. B 51 (1995) 12303-12312.

20. T. Frauenheim, G. Jungnickel, T. Köhler, P.K. Sitch, P. Blaudeck, Amorphous carbon: state of the art, World Scientific, Singapore, 1998, pp. 59-72.

21. P. Hohenberg, W. Kohn, Inhomogeneous electron gas, Phys. Rev. 136 (1964) B864-B871.

22. W. Kohn, P. Hohenberg, Self-consistent equations including exchange and correlation effects, Phys. Rev. 140 (1965) A1133-A1138.

23. J. M. Soler, E. Artacho, J.D. Gale, A. García, J. Junquera, P. Ordejón, D. Sánchez-Portal, The SIESTA method for ab initio order-N materials simulation, J. Phys. Condens. Matter 14 (2002) 2745-2779.

24. J.P. Perdew, K. Burke, M. Ernzerhof, Generalized Gradient Approximation Made Simple, Phys. Rev. Lett. 77 (1996) 3865-68.

25. K.R. Gschneidner Jr., Solid State Physics, Vol. 16, New York: Academic, p. 276.

26. R. Maezono, M.D. Towler, Y. Lee, R.J. Needs, Quantum Monte Carlo study of sodium, Phys. Rev. B 68 (2003) 1651031-9.

27. L. Schimka, J. Harl, G. Kresse, Improved hybrid functional for solids: The HSEsol functional, J. Chem. Phys. 134 (2011) 0241161-11.

28. N. Troullier, J.L. Martins, Efficient pseudopotentials for plane-wave calculations, Phys. Rev. B 43 (1991) 1993-2006.

29. S. Grimme, Semiempirical GGA-type density functional constructed with a long-range dispersion correction, J. Comp. Chem. 27 (2006) 1787-1799.

30. R. Nicklow, N. Wakabayashi, H. G. Smith, Lattice dynamics of pyrolytic graphite, Phys. Rev. B 5 (1972) 4951-4962.

31. K. Tasaki, Density Functional Theory Study on Structural and Energetic Characteristics of Graphite Intercalation Compounds, J. Phys. Chem. C 118 (2014) 1443-50.

32. H.J. Monkhorst, J.D. Pack, Special points for Brillouin-zone integrations, Phys. Rev. B 13 (1976) 5188-5192.

33. N. Yabuuchi, K. Kubota, M. Dahbi, S. Komaba, Research Development on Sodium-Ion Batteries, Chem. Rev. 114 (2014) 11636−11682.

34. J.B. Goodenough, Y. Kim, Challenges for Rechargeable Li Batteries, Chem. Mater. 22 (2010) 587-603.

35. D.R. McKenzie, Tetrahedral bonding in amorphous carbon, Rep. Prog. Phys. 59 (1996) 1611-1664.

36. Th. Frauenheim, G. Jungnickel, Th. Köhler, U. Stephan, Structure and electronic properties of amorphous carbon: from semimetallic to insulating behavior, Journal of Non-Crystalline Solids 182 (1995) 186-197.

37. G.A.F. Seber, Multivariate Observations, John Wiley & Sons, Hoboken, NJ, 1984.

38. H. Spath, Cluster Dissection and Analysis: Theory, FORTRAN Programs, Examples, Halsted Press, New York, 1985.

39. E. Lee, K. Persson, Li Absorption and Intercalation in Single Layer Graphene and Few Layer Graphene by First Principles, Nano Letters 12 (2012) 4624-4628.

40. Y. Liu, V.I. Artyukhov, M. Liu, A.R. Harutyunyan, B.I. Yakobson, Feasibility of Lithium Storage on Graphene and Its Derivatives, J. Phys. Chem. Lett. 4 (2013) 1737-1742.

41. F. Legrain, O.I. Malyi, S. Manzhos, Comparative ab initio study of lithium storage in amorphous and crystalline $TiO_2$, Proceedings of the 14[th] Asian Conference on Solid State Ionics, Singapore, 2014.

42. S. Komaba, W. Murata, T. Ishikawa, N. Yabuuchi, T. Ozeki, T. Nakayama, A. Ogata, K. Gotoh, K. Fujiwara, Electrochemical Na insertion and solid electrolyte interphase for hard-carbon electrodes and application to Na-ion batteries, Adv. Funct. Mater. 21 (2011) 3859-3867.
43. M.K. Aydinol, A.F. Kohan, G. Ceder, K. Cho, J. Joannopoulos, Ab initio study of lithium intercalation in metal oxides and metal dichalcogenides, Phys. Rev. B 56 (1997) 1354-65.
44. K. Momma, F. Izumi, VESTA 3 for three-dimensional visualization of crystal, volumetric and morphology data, J. Appl. Cryst. 44 (2011) 1272-1276.

# 7. Tables

Table 1. Defect formation energies $E_f$ of Li and Na in graphite and in *a*-C. Zero corresponds to the cohesive energy of Li and Na, respectively.

| host | Li $E_f$, eV | Na $E_f$, eV |
|---|---|---|
| Graphite | -0.085 | 0.760 |
| *a*-C 1 | -1.563 | -1.234 |
|  | -1.516 | -1.106 |
|  | -1.196 | 0.701 |
|  | -0.507 | 1.232 |
|  | -0.306 | 1.313 |
|  | -0.197 | 1.568 |
| *a*-C 2 | -1.539 | -0.891 |
|  | -1.053 | -0.845 |
|  | -0.697 | -0.153 |
|  | -0.613 | 0.407 |
|  | -0.543 | 1.247 |
|  | -0.137 | 1.666 |
|  | 0.285 |  |

## 8. Figure captions

Figure 1. Structures 1 (a) and 2 (b) of *a*-C (brown) with Li/Na insertion sites (green), and their radial distribution functions compared to experiment [19] (c). Visualization by VESTA [44]. The lowest energy insertion site in graphite is also shown in panel (d). Li sites are shown, and Na sites are visually similar.

Figure 2. (a) Filled symbols: the defect formation energies $E_f$ of Li and Na with respect to bulk Li and Na, respectively, in *a*-C. Circles and rhombuses are for the two amorphous structures. The corresponding $E_f$ values in graphite are also shown as empty black circles. (b) Dependence of $E_f$ on the coordination number $N$.

Figure 3. The defect formation energies per dopant atom ($E_f$) of Li/Na with respect to bulk Li/Na in *a*-C against Li/Na concentration $x$ ($x$ being the number of Li/Na metal atom per C atom). Circles and rhombuses are for the two amorphous structures.

## 9. Figures

Figure 1.

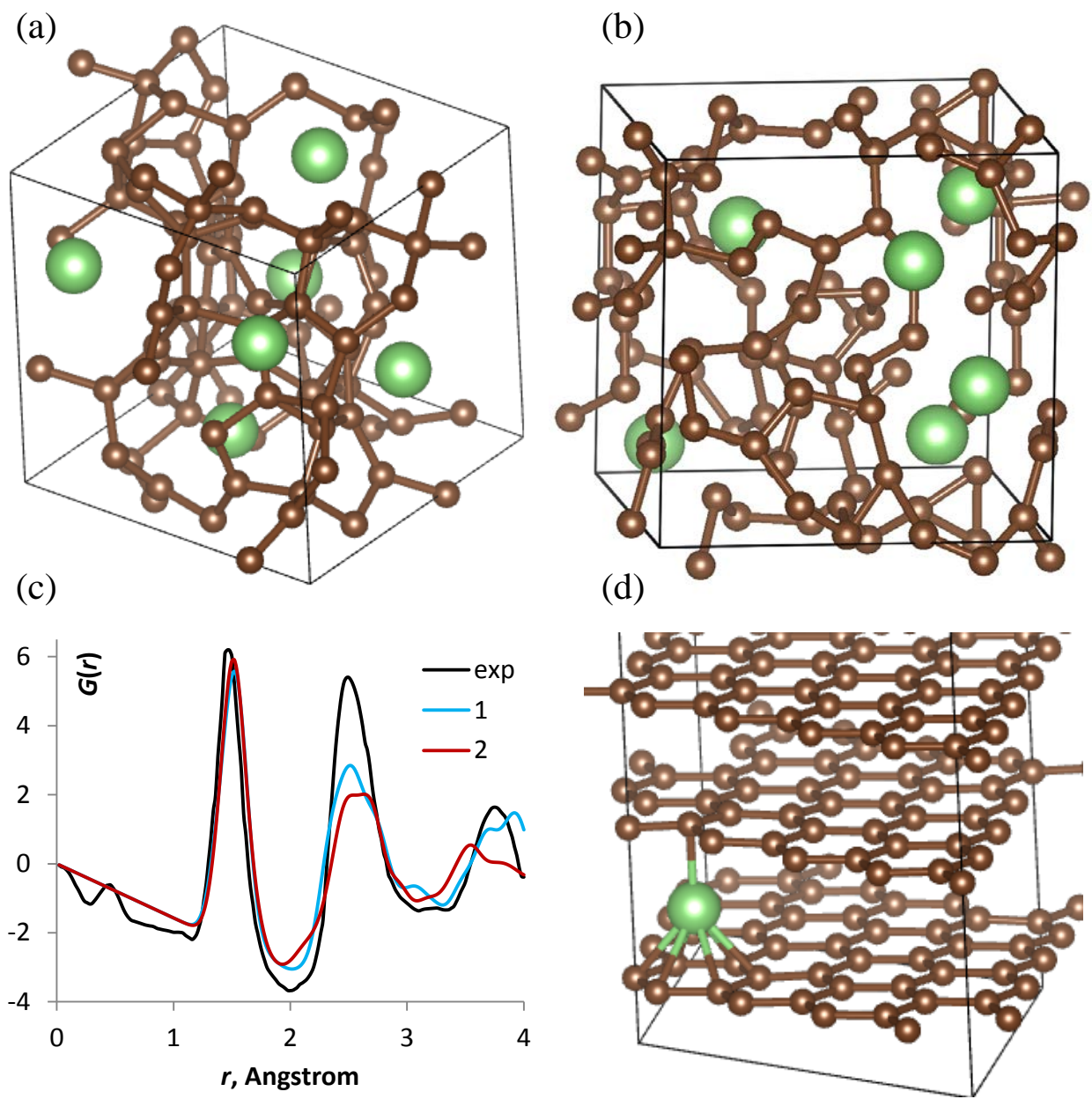

Figure 2.

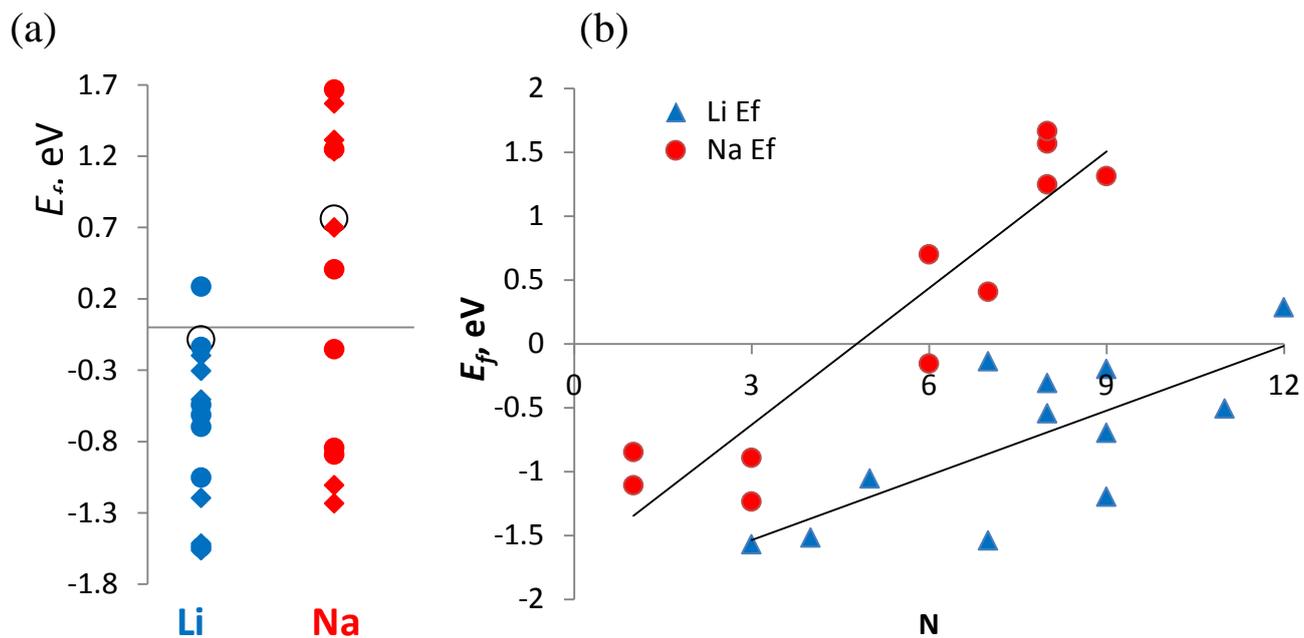

Figure 3.

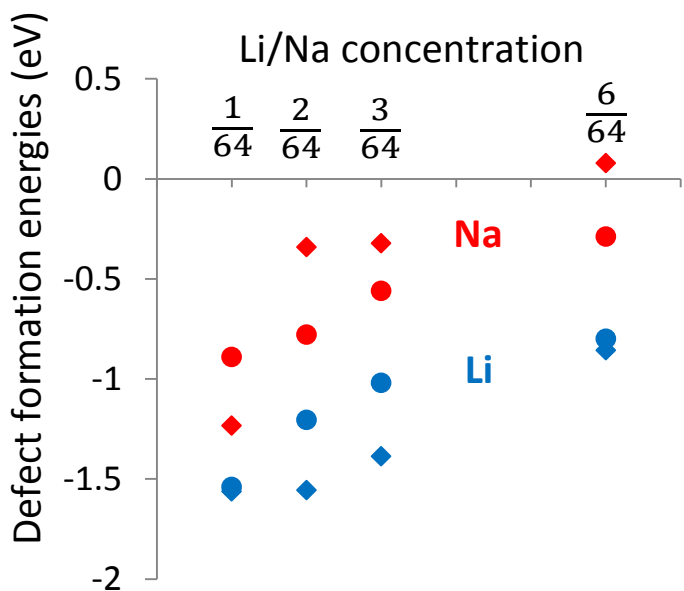